# Inverse-designed broadband low-loss grating coupler on thick lithium-niobate-on-insulator platform

Yijun Xie,[1,2] Mingming Nie,[1] and Shu-Wei Huang[1,*]

[1]*Department of Electrical, Computer, and Energy Engineering, University of Colorado Boulder, 425 UCB, Boulder, Colorado 80309, USA*
[2]*e-mail: yijun.xie@colorado.edu*
*\*shuwei.huang@colorado.edu*

**Abstract:** A grating coupler on 700-nm-thick Z-cut lithium-niobate-on-insulator platform with high coupling efficiency, large bandwidth, and high fabrication tolerance is designed and optimized by inverse design method. The optimized grating coupler is fabricated with a single set of e-beam lithography and etching process, and it is experimentally characterized to possess peak coupling efficiency of -3.8 dB at 1574.93 nm, 1-dB bandwidth of 71.7 nm, and 3-dB bandwidth of over 120 nm.

## 1. Introduction

Lithium niobate (LN) is a widely used material in different domains due to its large refractive indices, large transparency window, high second ($\chi^{(2)}$) and third ($\chi^{(3)}$) order nonlinearities as well as its excellent electro-optic (EO) property. The recently developed lithium niobate on insulator (LNOI) platform has brought new vigor and vitality to integrated photonics, which creates more opportunities for better performance and lower power consumption in various applications such as atomic clock, frequency synthesizer, LIDAR, and OCT-based bio-imaging [1]. These integrated devices are mainly demonstrated on thick LNOI platform with thickness ranging from 600 nm to 800 nm, where modes can be strongly confined and dispersion engineering can be flexibly achieved [2-9].

As for real-world applications, input and output couplers that are compatible with such integrated LNOI devices can help further improve the performance in different ways. Couplers with high coupling efficiency are essential to deliver high on-chip power for efficient nonlinear applications such as second harmonic generation (SHG) and electro-optic (EO) modulation [2-4]. Besides, high-efficient couplers are critical for chip-scale quantum applications demanding low loss. In addition, broadband couplers provide capability of retaining the spectral information and the accommodation of tunability for those broadband applications such as supercontinuum generation, femtosecond pulse generators, quadratic/Kerr combs/solitons generation and tunable lasers [5-9], . Therefore, couplers on such thick LNOI platform with not only high coupling efficiency but also large bandwidth are fundamentally important to fulfill the potential of those integrated LNOI devices.

Compared to the edge coupler requiring end-facet dicing, polishing and additional customized lensed fiber, grating coupler (GC) can typically provide high coupling efficiency with large placement flexibility. More importantly, GCs can be used along with fiber arrays which is convenient for multi-device testing and operation and thereby offers compelling advantages for high-volume production. In order to achieve high coupling efficiency, deep etching is usually required for GCs to increase the contrast between grating and trench regions, which is intrinsically difficult for LN. Conventionally, GCs with uniform periodicity and filling factor of the grating structures can only provide limited coupling efficiency with narrow bandwidth. Although lots of efforts have been made with manipulating the over cladding and forward design method where linear apodization and chirping is introduced and tuned [10-19], the demonstrated GCs cannot simultaneously provide high coupling efficiency and large bandwidth even with the compatibility with those photonic devices mentioned above sacrificed. The best GC that has been experimentally demonstrated on thick LNOI platform via direct LN etch so far exhibits only -6.3 dB peak coupling efficiency with 1-dB bandwidth and 3-dB bandwidth of 40 nm and 90 nm, respectively [20].

In this paper, we design and optimize the performance of the GC by inverse design method. In contrast to forward design method where only limited parameter space is explored to provide provisional optimization, inverse design method based on the gradient descent algorithm comprehensively optimizes the figure of merit (FOM), which is usually defined as the overall coupling efficiency in a given bandwidth, by updating the structure iteratively according to the gradient of FOM in much expanded parametric space. Our final design is based on a 700 nm-thick Z-cut LNOI platform with -2.94 dB peak coupling efficiency at 1572.4 nm, 1-dB bandwidth of 69 nm, and 3-dB bandwidth of 113 nm from the fundamental transverse electric (TE) mode input of a single mode fiber at telecom band (SMF-28), which has comprehensive improvement compared to the forward design. The proposed GC is subsequently fabricated on a 700-nm-thick z-cut LNOI chip and experimentally characterized with -3.8 dB peak coupling efficiency at 1574.9 nm, 1-dB bandwidth of 71.7 nm, and 3-dB bandwidth of over 120 nm, which agrees well with the designed performance. We believe the proposed GC that is perfectly compatible with those integrated devices can benefit various emergent applications.

## 2. Design and Optimization

We start the initial optimization with sweeping apodization and filling factors, after which initial design is found for further optimization, and the initial design is further optimized using inverse design method by iteratively adjusting the widths of each grating pillar and trench independently. The length of the grating pillars $l$ is fixed at 12 μm in order to match the mode from SMF-28 along the out-of-plane direction ($y$). Due to the large aspect ratio of the grating pillars, they can be considered as infinitely long slab waveguide along $y$ direction so that the simulation result from three-dimension FDTD (3D FDTD) method (Lumerical, Ansys) is almost identical to the result given by two-dimension FDTD (2D FDTD) method. Therefore, 2D FDTD method is chosen for all the simulations throughout the design and optimization section in the consideration of time and computational resources saving.

Some general parameters of the GC are kept as constants throughout the design section (unless otherwise stated): the number of periods is chosen to be 15 and the last period ends with a straight waveguide with width of 6 μm, the total thickness of LN $h$ and the etched depth $e$ are 700 nm and 450 nm, respectively. The base angle of the grating along the light propagation direction ($x$) is set to be 70 degrees based on our estimation from the SEM measurement while the base angle along $y$ direction is neglected. SMF-28 is located always at the center of the GC along $y$ direction and around 2.5 μm above the top surface of the GC along $z$ direction.

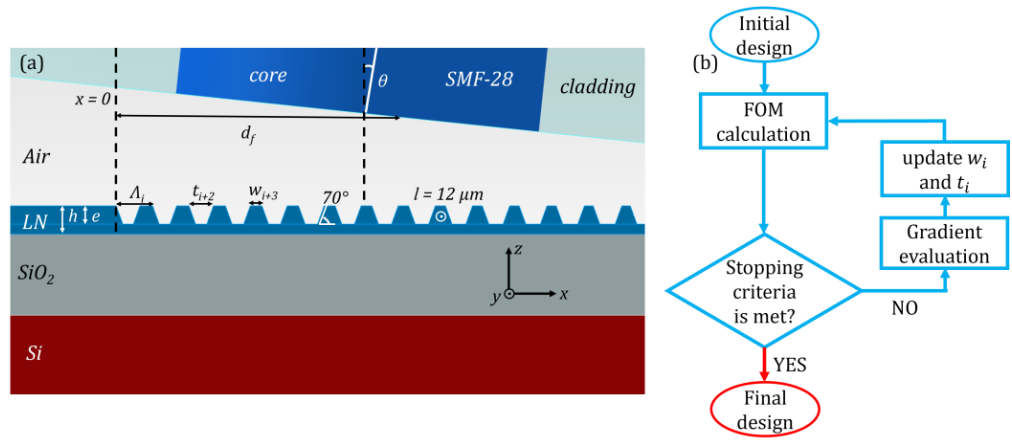

Fig. 1. (a) 2D Schematics of the grating coupler simulation, (b) flow chart of inverse design method.

The schematic of the simulation is shown in figure 1 (a). In order to obtain a good initial design for further optimization, we introduce linear apodization and chirping to the GC whose periods ($\Lambda_i$) and filling factors ($F_i$) of the gratings are given by [Eq. (1)] and [Eq. (2)]:

$$\Lambda_i = \frac{\lambda_c}{a+b\times F_i}, \tag{1}$$

$$F_{i+1} = F_1 - R \times x_i, \tag{2}$$

in which $F_1$ is the filling factor of the first grating period that is fixed at 0.35, $R$ is the apodization factor that is fixed at 0.01/μm, $x_i$ is the end of the current grating, $\lambda_c$ is the central wavelength that is set to be 1575 nm. For each period, the GC starts at $x = 0$ with the etched trench followed by the grating pillar and the width of the trench $t_i$ and the grating $w_i$ are given by $\Lambda_i - w_i$ and $\Lambda_i \times F_i$, respectively. The initial optimization is conducted by sweeping unitless parameters $a$ and $b$ with the sweep of the $x$ position of the SMF-28 $d_f$ and the angle $\theta$ of the SMF-28 with respect to normal direction of the GC ($z$) for each pair of $a$ and $b$. The optimized design is selected with peak coupling efficiency -4.48 dB at central wavelength 1571 nm, 1-dB and 3-dB bandwidths of 34 nm and 84 nm, respectively, and the GC parameters are listed in table 1.

**Table 1. Parameters of the optimized GC with linear apodization and chirping and the SMF-28**

| $F_0$ | $a$ | $b$ | $R$ ($\mu m^{-1}$) | $d_f$ ($\mu m$) | $\theta$ (deg) |
|---|---|---|---|---|---|
| 0.35 | 1.75 | 0.02 | 0.01 | 7 | 6 |

For further optimization, GC illustrated above is selected to be the initial design and full degrees of freedom on the width of each grating pillar $w_i$ and trench $t_i$ are granted while the $x$ position $d_f$ and the angle of the fiber $\theta$, the length of the gratings $l$, and the etched depth $e$ remain fixed at 7 μm, 6 degrees, 12 μm, and 450 nm, respectively. FOM is defined as the negative value of the average coupling efficiency in the wavelength range of 1555 nm to 1595 nm as expressed in [Eq. (3)]:

$$FOM = -\frac{\sum_{i=1}^{N} CE_i}{N}, \tag{3}$$

where $CE_i$ stands for the coupling efficiency at the reference wavelength. N is the number of reference wavelengths, which is set to be 51 so that the transmission data is recorded every 0.8 nm. Moreover, the constraint of the minimal feature size of 150 nm is added to avoid any super small structure that will impose excessive challenge on fabrication, and the stopping criteria is set to be FOM equals -0.5. To start with, FOM of the initial design is calculated by 2D FDTD simulation. Then the gradient of FOM with respect to each grating pillar and trench will be evaluated by adjoint method, which allows fast gradient evaluation regardless of number of parameters by only two FDTD simulations, if the stopping criteria is not satisfied. All the widths will be then updated accordingly by minimizer and FOM of the updated structure will be calculated iteratively. Eventually the iteration will end once the stopping criteria is met or minimal FOM is found, and the updated widths of each grating pillar and trench will be provided along with the optimal FOM as illustrated in figure 1 (b). The widths of gratings and trenches of the optimized GC are given in table 2.

By sweeping the x position $d_f$ and the angle $\theta$ of the SMF-28, the optimal position of the SMF-28 is subsequently found to be 6 μm away from the onset of the first trench at 8 degrees, and the performance of -2.94 dB peak coupling efficiency at 1572.4 nm, 1-dB bandwidth of 69 nm, and 3-dB bandwidth of 113 nm from the fundamental TE mode input of the SMF-28 is predicted. As listed in table 2, there is no super small structure determined by the constraint or any structure with irregular shape existing in the proposed GC.

**Table 2. Parameters of The Optimized GC by Inverse Design**

| | $t_1$ | $t_2$ | $t_3$ | $t_4$ | $t_5$ | $t_6$ | $t_7$ | $t_8$ |
|---|---|---|---|---|---|---|---|---|
| Trench size (nm) | 285 | 542 | 859 | 726 | 684 | 681 | 685 | 670 |
| | $t_9$ | $t_{10}$ | $t_{11}$ | $t_{12}$ | $t_{13}$ | $t_{14}$ | $t_{15}$ | |
| | 601 | 678 | 634 | 644 | 606 | 659 | 722 | |
| | $w_1$ | $w_2$ | $w_3$ | $w_4$ | $w_5$ | $w_6$ | $w_7$ | $w_8$ |
| Grating size (nm) | 302 | 157 | 161 | 230 | 272 | 248 | 257 | 295 |
| | $w_9$ | $w_{10}$ | $w_{11}$ | $w_{12}$ | $w_{13}$ | $w_{14}$ | $w_{15}$ | |
| | 253 | 280 | 252 | 210 | 196 | 175 | 6000 | |

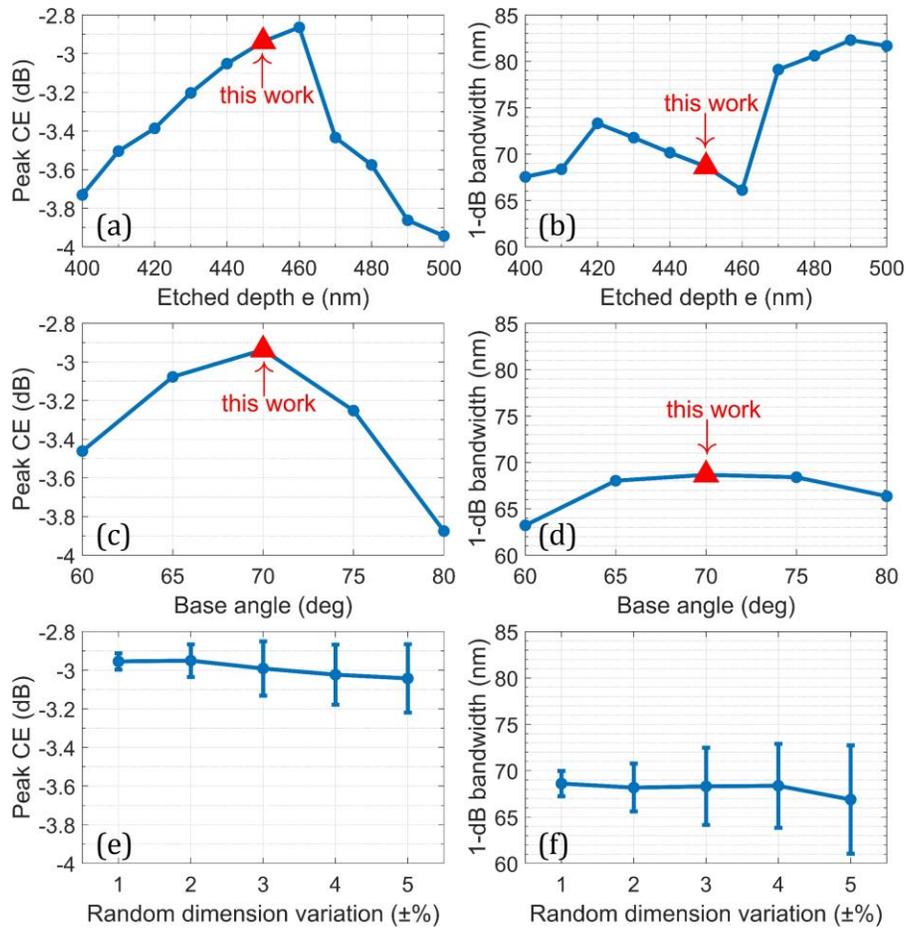

Fig. 2. The effect of different etched depth e on (a) peak coupling efficiency, and (b) 1-dB bandwidth; the effect of base angle on (c) peak coupling efficiency, and (d) 1-dB bandwidth; and the effect of random dimension variation on (e) peak coupling efficiency, and (f) 1-dB bandwidth.

In order to investigate the fabrication tolerance of the proposed GC, the variation of the peak coupling efficiency and 1-dB bandwidth are calculated with etched depth $e$ ranging from 400 nm to 500 nm and base angle fixed at 70 degrees, and with base angle varying from 60 degrees to 80 degrees and etched depth $e$ fixed at target value 450 nm in the consideration that different etching tools and recipes might render different base angles of the gratings. Furthermore, the random dimension variations from ±1% to ±5% from the designed values, which is common for E-beam lithography process, is also introduced to the size of each trench width $t_i$ and grating width $w_i$ separately while the etched depth $e$ and the base angle are both set to be the target values to evaluate how random dimension variations affect the performance of the optimized GC. The results are obtained from 100 individual simulations for each level of variation. The effects of different etched depths, base angles, and the random dimension variations on peak coupling efficiency, and 1-dB bandwidth are shown in figure 2 (a) and (b), figure 2 (c) and (d), and figure 2 (e) and (f), respectively.

As shown in figure 2 (a) and (b), the peak coupling efficiency of the GC varies from -3.9 dB to -2.85 dB while the 1-dB bandwidth varies from 66 nm to 82 nm in the exact opposite trend when the base angle is fixed at 70 degrees. The performance of the proposed GC remains consistent when the base angle of the gratings varies from 65 degrees to 75 degrees when the etched depth $e$ is kept at 450 nm as illustrated in figure 2 (c) and (d). However, the peak coupling efficiency starts to decay severely to -3.8 dB as the base angle approaches 80 degrees while the 1-dB bandwidth gets narrowed down by around 6 nm as the gratings get flatter.

On the other hand, the performance of the proposed GC is not prone to the random dimension variations as depicted in figure 2 (e) and (f). There is only about 0.3 dB peak coupling efficiency, and 8 nm 1-dB bandwidth reduction, respectively, for the most extreme cases where the random dimension variation is ±5% for the size of each trench and grating pillar, which indicates the overall high fabrication tolerance of the proposed GC.

### 3. Fabrication and Measurement

The proposed GC is fabricated on a 1 cm$^2$ 700-nm-thick Z-cut LNOI (NanoLN) chip. After thoroughly cleaning the chip, a 1.2-um-thick layer of AR-P 6200 (Allresist) is spin-coated and baked at 180 °C for 2 minutes. An E-beam lithography is conducted using JEOL JBX-6300 fs to define the soft mask and a subsequent argon ion milling is conducted for 12 minutes and 40 seconds in order to achieve the etched depth of 450 nm. After the residual mask is removed by remover PG, the LNOI chip is soaked in RCA solution (1:1:5 of $H_2O_2$, $NH_4OH$, and DI water) overnight at room temperature for the re-deposition removal to reduce the sidewall roughness. Each device consists of a 100-μm-long straight waveguide with the top width of 1.8 μm, two 300-μm-long tapers with the top width adiabatically expanding from 1.8 μm to 12 μm on both sides, and two proposed GCs. In order to compare the performance of the GC before and after the inverse design optimization, two GCs before the inverse design optimization with the exact same waveguide and tapers are also fabricated on the same chip. SEM images of the optimized GC by inverse design method are shown in figure 3.

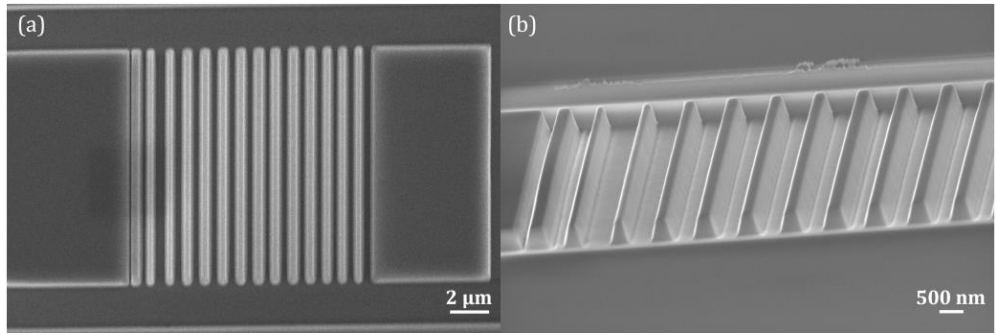

Fig. 3. SEM images of the fabricated GC (a) Top view, (b) Side view.

For the coupling efficiency characterization, two SMF-28s are cleaved and clamped by fiber holders that are mounted on rotational stages so that the fibers can be adjusted to the desired angle with respect to the normal direction of the GCs. Rotational stages are subsequently fixed onto 3-axis translation stages, which are placed on the opposite side of the fabricated device. A polarization controller connects the input SMF-28 and the tunable laser input from Santec TSL-710 while the output SMF-28 is connected to a photodetector whose output signal is collected by a digitizer (CSE1222 Razor Express, Gage) with 100Ks/s sampling rate for the data acquisition. Santec is continuously scanned from 1500 nm to 1640 nm at 100nm/s scan rate with 5 dBm nominal optical output power. The acquired data is converted to optical power by electrical-optical signal conversion and calibration and then the coupling efficiency of the GC can be further extracted. The simulated and the measured coupling efficiency after moving average process of the GC with linear apodization and chirping, and the GC optimized by inverse design are illustrated in figure 4 (a) and 4 (b), respectively.

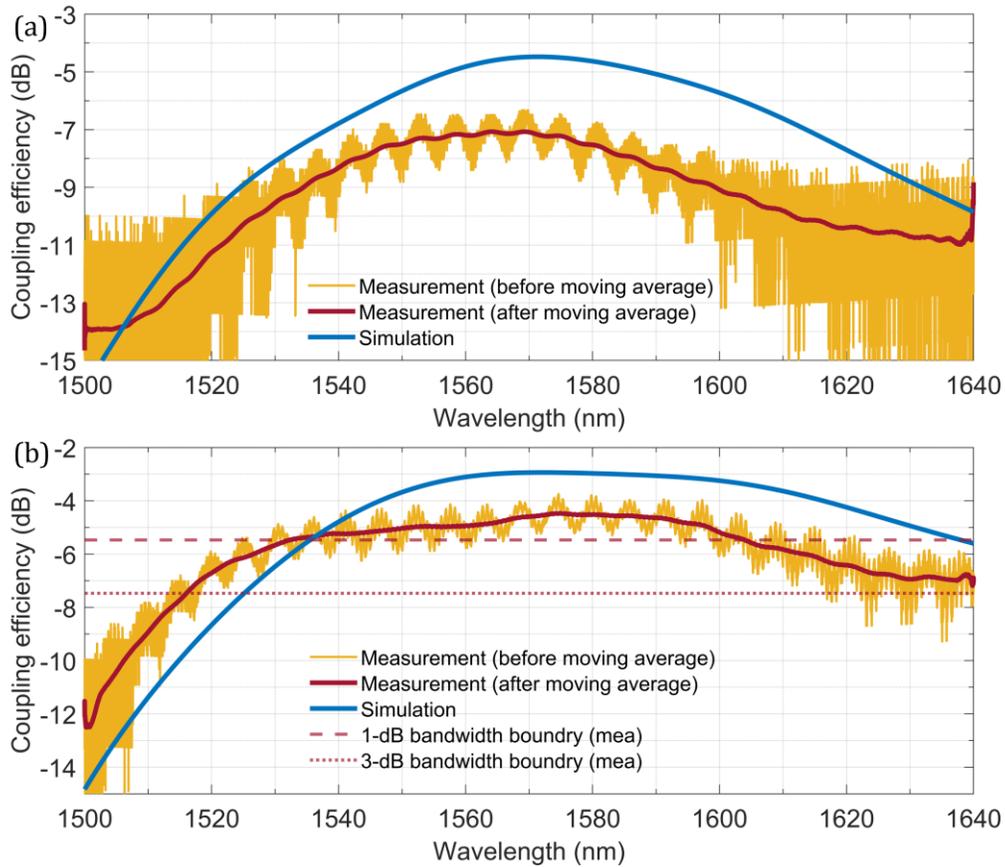

Fig. 4. Simulated and measured coupling efficiency of (a) the GC with linear apodization and chirping, (b) the GC optimized by inverse design method.

As shown in figure 4(a), the GC with linear apodization and chirping has similar performance compared to the simulation results with 1-dB bandwidth of 45.7 nm and 3-dB bandwidth of 85.8 nm except for the peak coupling efficiency. The measured peak coupling efficiency is -6.33 dB (-7.1±0.77 dB from moving average process) at 1569.3 nm with almost -2 dB difference from the simulation result, and we attribute this discrepancy mainly to the fabrication imperfection due to the proximity of this device to the edge of the chip. Compared to the results with linear apodization and chirping, the performance of the GC has comprehensively improved after optimization with the peak coupling efficiency soared from -6.33 dB to -3.8 dB (-4.5±0.7 dB from moving average process), 1-dB bandwidth expanded from 45.7 nm to 71.7 nm, and 3-dB bandwidth from 85.8 nm to over 120 nm, respectively, as plotted in figure 4(b). Besides, the measured results of the GC by inverse design match well with the simulation results with even slightly larger bandwidths but lower peak coupling efficiency, which can be explained by the random dimension variations of grating structures as discussed at the end design and optimization section.

## 4. Conclusion and outlook

In summary, a GC on 700-nm-thick z-cut LNOI with high coupling efficiency and large bandwidth is proposed and optimized by inverse design method, and the optimized GC is fabricated with only a single set of E-beam lithography and etching process. Comparable performance predicted by the simulation of peak coupling efficiency -3.8 dB at central wavelength 1574.9 nm, 1-dB bandwidth 71.7 nm, and 3-dB bandwidth over 120 nm is experimentally demonstrated. Comprehensive improvement has been observed in both peak coupling efficiency and bandwidths compared to the initial GC design with linear apodization and chirping.

This is the first proposed GC design optimized by inverse design method on LNOI platform as well as the first experimental demonstration of GC on the thick LNOI platform that exhibits both high coupling loss and large 1-dB and 3-dB bandwidth at the same time. Moreover, the proposed GC has high fabrication tolerance without any additional over cladding patterning or the need of bottom reflector. In principle, the GC with improved performance on LNOI platform with different crystalline orientations, total thicknesses, and etched depths and even on all other platforms can also be achieved following the same optimization strategies. One can also target at maximizing bandwidth or manipulating central wavelength by customizing FOM for future research.

**Funding.** NSF OMA 2016244; ONR N00014-22-1-2224.

**Acknowledgements.** Y. Xie thanks L. Rukh, C. Tang, and V. Babicheva for fruitful discussions and A. R. James, J. Nogan, and D. Webb for process advice and N. Prakash, J. Musgrave, J. Bartos for the help. This work was performed, in part, at the Center for Integrated Nanotechnologies, an Office of Science User Facility operated for the U.S. Department of Energy (DOE) Office of Science by Los Alamos National Laboratory (Contract 89233218CNA000001) and Sandia National Laboratories (Contract DE-NA-0003525).

**Disclosures.** Authors declare no conflicts of interest.

**Data availability.** Data underlying the results presented in this paper are not publicly available at this time but may be obtained from the authors upon reasonable request.

## References

1. M. Nie, Y. Xie, B. Li, *et al*., "Photonic frequency microcombs based on dissipative Kerr and quadratic cavity solitons," Progress in Quantum Electronics **100437** (2022).
2. J.-Y. Chen, C. Tang, Z.-H. Ma, *et al*., "Efficient and highly tunable second-harmonic generation in Z-cut periodically poled lithium niobate nanowaveguides," Opt. Lett. **45**(13), 3789–3792 (2020).
3. J. Lu, J. B. Surya. X. Liu, *et al*., "Periodically poled thin-film lithium niobate microring resonators with a second-harmonic generation efficiency of 250,000%/W," Optica **6**(12), 1455–1460 (2019).
4. C. Wang, M. Zhang, X. Chen, *et al*., "Integrated lithium niobate electro-optic modulators operating at CMOS-compatible voltages," Nature **562**(7725), 101–104 (2018).
5. M. Yu, B. Desiatov, Y. Okawachi, *et al*., "Coherent two-octave-spanning supercontinuum generation in lithium-niobate waveguides," Opt. Lett. **44**(5), 1222–1225 (2019).
6. M. Yu, D. Barton ⦀, R. Cheng, *et al*., "Integrated femtosecond pulse generator on thin-film lithium niobate," Nature **612**(7939), 252–258 (2022).
7. Y. He, Q.-F. Yang, J. Ling, *et al*., "Self-starting bi-chromatic LiNbO$_3$ soliton microcomb," Optica **6**(9), 1138–1144 (2019).

8. M. Nie, and S.-W. Huang, "Quadratic soliton mode-locked degenerate optical parametric oscillator," Opt. Lett. **45**(8), 2311–2314 (2020).
9. L. Ledezma, A. Roy, L. Costa, *et al.*, "Octave-spanning tunable infrared parametric oscillators in nanophotonics," Sci. Advances **9**(30), eadf9711 (2023).
10. J. Jian, P. Xu, H. Chen, *et al.*, "High-efficiency hybrid amorphous silicon grating couplers for sub-micron-sized lithium niobate waveguides," Opt. Express **26**(23), 29651–29658 (2018).
11. Y. Li, X. Huang, Z. Li, *et al.*, "Efficient grating couplers on a thin film lithium niobate–silicon rich nitride hybrid platform," Opt. Lett. **45**(24), 6847–6850 (2020).
12. X. Ma, C. Zhuang, R. Zeng, *et al.*, "Polarization-independent one-dimensional grating coupler design on hybrid silicon/LNOI platform," Opt. Express **28**(11), 17113–17121 (2020).
13. Z. Ruan, J. Hu, Y. Xue, *et al.*, "Metal based grating coupler on a thin-film lithium niobate waveguide," Opt. Express **28**(24), 35615–35621 (2020).
14. B. Chen, Z. Ruan, K. Chen, *et al.*, "One-dimensional grating coupler on lithium-niobate-on-insulator for high-efficiency and polarization-independent coupling," Opt. Lett. **48**(6), 1434–1437 (2023).
15. X. Zhou, Y. Xue, F. Ye, *et al.*, "High coupling efficiency waveguide grating couplers on lithium niobate," Opt. Lett. **48**(12), 3267–3270 (2023).
16. I. Krasnokutska, R. J. Chapman, J.-L. J. Tambasco, *et al.*, "High coupling efficiency grating couplers on lithium niobate on insulator," Opt. Express **27**(13), 17681–17685 (2019).
17. S. Kang, R. Zhang, Z. Hao, *et al.*, "High-efficiency chirped grating couplers on lithium niobate on insulator," Opt. Lett. **45**(24), 6651–6654 (2020).
18. B. Chen, Z. Ruan, X. Fan, *et al.*, "Low-loss fiber grating coupler on thin film lithium niobate platform," APL Photonics **7**(7), 076103 (2022).
19. E. Lomonte, F. Lenzini, and W. H. P. Pernice, "Efficient self-imaging grating couplers on a lithium-niobate-on-insulator platform at near-visible and telecom wavelengths," Opt. Express **29**(13), 20205–20216 (2021).
20. S. Yang, Y. Li, J. Xu, *et al.*, "Low loss ridge-waveguide grating couplers in lithium niobate on insulator," Opt. Materials Express **11**(5), 1366–1376 (2021).